\documentclass{aa}

\usepackage{graphicx}
\usepackage{txfonts}
\usepackage{hyperref}
\usepackage{amsmath}

\begin{document}

\title{Satellite stellar-to-halo mass relations in primordial black hole universes}
\titlerunning{Satellite stellar-to-halo relations in PBH universes}

\author{
Patricio Colazo\inst{\ref{Iate},\ref{OAC}}\and
Nelson Padilla\inst{\ref{Iate},\ref{OAC}} \and
Federico Stasyszyn\inst{\ref{Iate},\ref{OAC}} 
}  

\institute{
CONICET. Instituto de Astronomía Teórica y Experimental (IATE). Laprida 854, Córdoba X5000BGR, Argentina \label{Iate}\\
\email{nelson.padilla@unc.edu.ar}
\and 
Universidad Nacional de Córdoba (UNC). Observatorio Astronómico de Córdoba (OAC). Laprida 854, Córdoba X5000BGR, Argentina\label{OAC}
}

\date{\today}

\abstract
{Primordial black holes (PBHs), in particular those following a
fixed-conformal-time (FCT) extended mass function, can modify small-scale
structure through enhanced primordial power and discreteness-driven Poisson
fluctuations, altering both subhalo abundances and internal structure.}
{We study the effects that these alterations induce on satellite
stellar-to-halo mappings, bound-remnant compactness, and the possible
connection to compact dark perturbers.}
{We compare two dark-matter-only cosmological simulations, a
fiducial CDM run and an FCT/PBH run. Since these simulations do not form
galaxies, we construct an approximate abundance-based stellar-to-halo mapping
using the progenitor virial mass at last isolation. The CDM mapping is anchored
to a redshift-dependent abundance-matching stellar-to-halo relation, and the FCT mapping is derived from the relative cumulative subhalo abundances. We use resolved SOAP
\(V_{\max}\) and \(R_{\max}\) measurements to characterize present-day subhalo
compactness and supplement them with a subgrid estimate of mass segregation
in the unresolved PBH population.}
{FCT produces a mass-dependent enhancement of the subhalo abundance,
but abundance-based stellar-mass reassignment makes satellite stellar-mass
functions only weak discriminants between the models. Internal structure
provides a stronger test: FCT subhaloes are more compact and have larger
NFW-reconstructed enclosed masses at fixed radius, strengthening a
Too-Big-To-Fail-like tension if all compact subhaloes are assumed to host
luminous satellites. FCT subhaloes also form earlier, accounting for much of
the low-mass compactness difference, while a subgrid estimate of PBH mass
segregation produces an additional central perturbation with typical
\(\mu\simeq0.4\)--0.5. Neither effect fully accounts for the resolved FCT
compactness enhancement.}
{Changes in the galaxy--halo mapping can therefore partly hide the
enhanced FCT subhalo abundance in satellite counts, whereas internal
dynamical constraints remain more discriminating. Satellite abundances,
satellite dynamics, and compact dark perturbers provide complementary probes
of PBH-induced small-scale structure.}
{}

\maketitle

\section{Introduction}
\label{sec:introduction}

The nature of dark matter remains one of the central open problems in
cosmology. Among the many proposed candidates, primordial black holes (PBHs)
are especially interesting because they can arise from enhanced primordial
fluctuations in the early Universe and can leave signatures on structure
formation across a wide range of scales
\citep{Inomata_2017,Clesse_2015,Carr_2021,Sasaki_2018}. Depending on the
formation scenario, PBHs may follow either a horizon-crossing (HC) or
fixed-conformal-time (FCT) mass function \citep{Sureda_2021}. In the FCT
case, PBHs form nearly simultaneously during the radiation era, producing an
extended mass function whose observational constraints can differ
substantially from those obtained for monochromatic or narrow log-normal PBH
populations.

PBHs can affect structure formation in two related ways. First, the
inflationary models that produce PBHs typically require enhanced small-scale
primordial power. Second, the discreteness of the PBH population introduces a
Poisson contribution to the matter power spectrum, modifying the growth of
fluctuations on small scales \citep{Padilla_2021}. These effects can change
the abundance of low-mass haloes and subhaloes relative to standard
\(\Lambda\)CDM
\citep{Inman_2019,Colazo_II,Zhang__2024,Matteri,Liu_2022,Colazo_2025}.
This is qualitatively different from warm or fuzzy dark matter models, which
usually suppress low-mass structure through a cutoff in the initial power
spectrum
\citep{colin_substructure_2000,zentner_halo_2003,he_extending_2023,elgamal_no_2024}.

Most previous work has focused on the direct detectability of modified
substructure abundances, for example through strong gravitational lensing,
flux-ratio anomalies, stellar streams, microlensing, or other gravitational
probes
\citep{dalal_direct_2002,Vegetti_2009,li_2016,Enzi_2021,nightingale_2023,xiao_detecting_2024,weinberg_cold_2015}.
These observables are especially relevant for dark subhaloes, commonly
associated with masses below \(\sim10^9\,M_\odot\), where galaxy formation is
inefficient and the objects may be invisible except through their
gravitational effects
\citep{Mena_2019,Gow_2020,Villanueva_2021,Ziparo_2022,Byrnes_2024}.

This leaves open the complementary question of how PBH-like physics would
affect the galaxy--halo connection in the luminous satellite population.
The discreteness of a PBH population generates an additional Poisson
isocurvature contribution to the matter power spectrum, which can enhance
the formation of low-mass haloes at high redshift
\citep{Afshordi2003,Gong2017}.  Early cosmological N-body calculations
subsequently showed explicitly that this contribution can accelerate the
nonlinear formation of the first bound structures in PBH dark-matter
cosmologies \citep{Inman_2019}.  More recent hydrodynamical and
dark-matter-only simulations have investigated the consequences for the
first stars, high-redshift galaxies, and halo abundances
\citep[e.g.][]{Liu_2022,boyuan_2022_stars,Colazo_II,Colazo_2025}.  Here we
address a complementary low-redshift question.  If PBH/FCT physics alters the
subhalo population of Milky-Way-mass haloes, it should also change the
stellar-to-halo mass mapping inferred for satellite galaxies.

The Milky Way satellite population provides a useful low-redshift laboratory
for this question. Satellite abundances constrain the connection between
subhaloes and stellar mass, while half-light radii and stellar velocity
dispersions probe the enclosed dynamical masses of the host subhaloes
\citep[e.g.][]{McConnachie2012,DrlicaWagner2020,Wolf2010,Pace2024LVDB}.
These diagnostics are complementary: changes in subhalo abundance may be
partly absorbed into a modified stellar-to-halo mass mapping, whereas changes
in internal structure remain accessible through satellite dynamics.

Here we compare matched PBH/FCT and \(\Lambda\)CDM dark-matter-only
simulations to determine how the modified subhalo population affects the
satellite stellar-to-halo mass relation, the cumulative stellar-mass function,
and the internal masses of Milky Way satellites. We also examine more massive
group-scale hosts, where compact low-mass substructure is relevant for
gravitational lensing. Since individual PBHs are unresolved, we supplement
the resolved subhalo structure with a subgrid estimate of PBH mass
segregation and assess where the high-mass tail of the PBH distribution could
produce an additional inner mass perturbation.

Section~\ref{sec:simulations} describes the simulations, halo catalogues,
stellar-mass assignment, structural diagnostics, and PBH mass-segregation
estimate. Section~\ref{sec:modified_shmr} presents the abundance-corrected
stellar-to-halo mass relation. Section~\ref{sec:mw_satellites} compares the
predictions with Milky Way satellite stellar masses and dynamical masses.
Section~\ref{sec:beyond_mw} discusses the GAMA group comparison and compact
dark substructure, and Sect.~\ref{sec:discussion_conclusions} summarizes the
results.

\section{Simulations and subgrid model of PBH effects}
\label{sec:simulations}

We use the same simulation suite described in \citet{Colazo_2025}. The initial
conditions are built from a standard primordial power spectrum on large
scales,
\[
P_{\rm primordial}(k)=A_s\left(\frac{k}{k_0}\right)^{n_s},
\]
where \(k_0=0.05\,h{\rm Mpc}^{-1}\) and \(n_s\) is taken from the Planck
cosmology. For the non-standard models, the spectrum is modified above a
pivot scale \(k_{\rm piv}\simeq 10\,h{\rm Mpc}^{-1}\), where it changes to a
blue spectral index \(n_b\). In the PBH/FCT case, the matter power spectrum also includes the
Poisson contribution associated with the discrete nature of the PBH
population,
\[
P(k,z)=\widetilde{P}_{\rm primordial}(k)T^2(k)D_1^2(z)
+f_{\rm PBH}^2P_{\rm Poisson}^{\rm PBH}(k,z).
\]
Here \(P_{\rm Poisson}^{\rm PBH}(k,z)\) denotes the PBH discreteness
contribution to the matter power spectrum, including its scale- and
redshift-dependent evolution, calculated following the prescription of
\citet{Padilla_2021}; we refer to that work for its explicit
form.  The same prescription and initial power spectrum were used in
\citet{Colazo_2025}.  PBHs are not represented as a separate
particle species in the simulations; their effect is encoded solely
through the modified initial power spectrum.

The FCT simulation analysed here was generated with \(n_b=2.5\),
\(f_{\rm PBH}=1\), and characteristic PBH mass
\(M_\ast=10^2\,M_\odot h^{-1}\), following the fixed-conformal-time
mass-function framework of \citet{Sureda_2021}.  Individual PBHs are not
resolved as particles; their effect enters through the modified initial
matter power spectrum.

The initial conditions are generated at \(z=1200\) using
\textsc{MUSIC2-monofonIC} \citep{MUSIC2-MonophonIC}. This high starting redshift is required because
the FCT model has enhanced small-scale power, and the initial conditions must
remain in the quasi-linear regime over the resolved range of wavenumbers.
The simulations use a box size of \(35\,h^{-1}{\rm Mpc}\) with \(1024^3\)
dark matter particles of mass $\simeq 5 \times 10^{6}\,M_\odot$. The softening length is $1.8$ comoving kpc
and we impose a minimum halo resolution threshold of 30 dark matter particles.

The suite used here consists of only two of the simulations from
\citet{Colazo_2025}: a fiducial \(\Lambda\)CDM run, and a PBH/FCT run
including both the blue index and the Poisson contribution. Both simulations
were evolved with the \textsc{SWIFT} code, using the fiducial cosmological
parameters of \citet{Planck_2020}. We use 120 snapshots between $z=10$ and $z=0$ to follow the
evolution of haloes and subhaloes and to identify infall quantities.

As shown by \citet{Colazo_2025}, the additional small-scale power has little
effect on halo clustering on the resolved scales, while its main consequences
are an enhanced abundance of low-mass haloes and changes in their internal
structure. A corresponding prediction for galaxy clustering would additionally
require a model for galaxy formation and halo occupation, which is not included
in these dark-matter-only simulations.

\subsection{Halo catalogues and merger trees}
\label{sec:halo_catalogues}

Haloes and subhaloes are identified with \textsc{HBT-HERONS}
\citep{ForouharMoreno2025HBT,Han2018HBTplus}, a history-based descendant of
\textsc{HBT+} that follows self-bound structures across snapshots using
persistent particle membership and assigns each object a unique track
identifier. Infall is defined operationally as the HBT-HERONS
\texttt{SnapshotOfLastIsolation}, the last snapshot at which the object is
identified as an isolated central. This proxy need not coincide exactly with
the first crossing of the host virial radius, particularly for backsplash or
splashback objects.

We supplement the HBT-HERONS catalogue with spherical-overdensity and
aperture properties computed by \textsc{SOAP} \citep{McGibbon2025SOAP}. SOAP
uses the halo centres and particle memberships supplied by the input catalogue
to compute quantities including \(M_{200{\rm c}}\), \(R_{200{\rm c}}\), spin,
and NFW-equivalent concentration estimates. SOAP spherical-overdensity masses
include all particles enclosed within the corresponding radius, whereas HBT
bound masses include only particles gravitationally bound to the subhalo.
Accordingly, we use SOAP \(M_{200{\rm c}}\) for host selection and for the
isolated progenitor mass entering the stellar--halo mass assignment, while
HBT bound quantities are used to characterize surviving substructure and
tidal stripping.

For each satellite, we record the present-day bound mass \(M_{\rm now}\), the
bound mass at infall \(M_{\rm bound,infall}\), the maximum circular velocity
at infall \(V_{\max,{\rm infall}}\), and the peak value over the subhalo
history, \(V_{\max,{\rm peak}}\). Bound-substructure and stripping diagnostics use HBT bound masses:  
\(M_{\rm bound,infall}\) is adopted when the progenitor scale is required, since \(M_{\rm now}\) can be strongly reduced after accretion, whereas
\(M_{\rm now}\) is used for present-day substructure comparisons. The stellar-mass
assignment instead uses the SOAP spherical-overdensity mass of the progenitor
at last isolation, \(M_{200{\rm c},\rm sub}^{\rm infall}\), converted to the
virial mass convention of the adopted stellar--halo mass relation.
Although \(V_{\max,{\rm peak}}\) is retained in the catalogues, it is not used
as the primary ranking variable in this work.

We characterize the assembly epoch of each subhalo by its half-mass formation
redshift, \(z_{\rm form}\), defined as the redshift at which the main
progenitor first reaches one half of its peak bound mass. This quantity is
distinct from the infall redshift: \(z_{\rm form}\) describes the assembly of
the subhalo, whereas \(z_{\rm infall}\) marks its last snapshot as an isolated
central.

The structural quantities used for isolated progenitors and present-day
satellites are also defined differently. For progenitors at last isolation,
we use the NFW-equivalent spherical-overdensity concentrations estimated by
SOAP from the density-profile moment method. For stripped present-day
satellites, a formal \(c_{200{\rm c}}\) is generally not well defined, so we
use \(V_{\max,{\rm now}}\) and \(R_{\max,{\rm now}}\) as a bound-remnant
structural proxy. When an NFW profile is required for the satellite dynamical
comparison, we construct an approximate profile satisfying
\(R_{\max}=2.1626\,r_s\) and
\(M(<R_{\max})=V_{\max}^2R_{\max}/G\). This reconstruction is a diagnostic of
the resolved compactness and enclosed mass of the bound remnant, rather than
a direct measurement of the unresolved central density profile. The PBH
mass-segregation calculation introduced below is therefore a separate
subgrid interpretation, not a component resolved by the simulation.

\subsection{Subgrid PBH mass segregation in FCT substructures}
\label{sec:subgrid_pbh_occupation}

The FCT simulation does not contain individual PBHs as resolved particles.
PBHs enter only through the modified initial linear matter power spectrum,
including the enhanced small-scale primordial contribution and the Poisson
isocurvature term generated by PBH discreteness
\citep{Mack2007,Gong2017,Inman_2019,Hutsi2019}. The simulated subhalo
abundances and resolved \(V_{\max}\)--\(R_{\max}\) structure therefore
describe the nonlinear evolution of this coarse-grained density field, not
an object-by-object realization of PBHs within individual haloes.

We use a separate subgrid calculation to estimate whether the high-mass tail
of the same PBH mass function can subsequently segregate toward subhalo
centres through dynamical friction. This post-processing model addresses the
granular, object-level effect that is not resolved in the simulation. It does
not modify the simulated density field or model a fully discrete PBH halo,
stochastic relaxation, baryonic contraction, or the formation and tidal
evolution of PBH-centred minihaloes. Instead, it estimates the PBH mass that
can sink within the available time and compares it with the enclosed mass of
reference NFW profiles.

We assign the PBH mass function statistically to each FCT subhalo, taking its
infall bound mass \(M_{\rm bound,infall}\) as the available pre-accretion PBH
reservoir. For the FCT model, the low-mass slope is related to \(n_b\)
\citep{Sureda_2021} by
\[
\frac{{\rm d}n}{{\rm d}M_{\rm PBH}}
\propto
M_{\rm PBH}^{-(9-n_b)/6}.
\]
The distribution is normalized to the total PBH mass fraction over the
adopted mass range. Because the dynamical-friction time decreases with PBH
mass, only the high-mass tail contributes appreciably to the sunk component.

The sinking radius is estimated with the Chandrasekhar dynamical-friction
time for a massive object on a circular orbit,\footnote{Circular orbits are
assumed only to define a reproducible sinking criterion; PBHs in
collisionless haloes should have eccentric orbits, which can change the decay
time by factors of order unity.}
\[
t_{\rm df}(r,M_{\rm PBH})
\simeq
\frac{1.17\,V_{\rm c}(r)\,r^2}
     {G\,M_{\rm PBH}\ln\Lambda},
\qquad
V_{\rm c}^2(r)=\frac{G\,M(<r)}{r}.
\]
Here \(\ln\Lambda\) is the Coulomb logarithm, which parametrizes the ratio
between the largest and smallest effective impact parameters contributing to
the gravitational wake.  We adopt a fixed fiducial value
\(\ln\Lambda=10\), appropriate for an order-of-magnitude sinking criterion.

For each PBH mass, we define \(r_{\rm sink}(M_{\rm PBH})\) as the largest
pre-segregation radius within the subhalo from which a PBH can sink to the
centre within the available time,
\[
t_{\rm df}\!\left(r_{\rm sink},M_{\rm PBH}\right)
=
13.8\,{\rm Gyr}.
\]
Only PBHs located within this radius before segregation are counted as able
to sink by the present time.  Thus \(r_{\rm sink}\) sets the radial fraction
of PBHs of each mass that contributes to the sunk component.

Assuming that the pre-segregation PBH population follows a chosen reference
NFW profile independently of PBH mass, the contribution per
logarithmic PBH mass interval is
\[
{\cal I}_{\rm sunk}(M)
\equiv
\psi(M)\,
f_{\rm NFW}\!\left[r_{\rm sink}(M)\right],
\]
where \(M\equiv M_{\rm PBH}\). The expected sunk PBH mass is then
\[
M_{\rm sunk}
=
f_{\rm PBH} M_{\rm bound,infall}
\int {\cal I}_{\rm sunk}(M)\,{\rm d}\ln M .
\]

Here \(\psi(M)\) is the normalized PBH mass-fraction distribution per
logarithmic mass interval,
\[
{\rm d}f_{\rm PBH}(M)
=
f_{\rm PBH}\,\psi(M)\,{\rm d}\ln M,
\qquad
\int\psi(M)\,{\rm d}\ln M=1,
\]
over the adopted PBH mass range, whose lower limit is set by the evaporation
mass for a lifetime of 13.8 Gyr. The function \(f_{\rm NFW}(r)\) is the
fraction of the chosen reference-profile mass enclosed within \(r\), capped
at unity. We define \(r_{\rm eff}\) as the median
\(r_{\rm sink}(M)\), weighted by the sunk-mass integrand
\({\cal I}_{\rm sunk}(M)\).

We carry out this calculation for two reference profiles.  First, we use a
CDM-median reference profile at the same \(M_{\rm bound,infall}\), obtaining
\(M_{\rm sunk}^{\rm CDMref}\) and \(r_{\rm eff}^{\rm CDMref}\).  This defines
\[
\mu_{\rm CDMref}
=
\frac{M_{\rm sunk}^{\rm CDMref}}
     {M_{\rm NFW,CDMref}(<r_{\rm eff}^{\rm CDMref})}.
\]
This comparison asks whether PBH sinking from an initially CDM-like profile
can generate an order-unity central mass perturbation, beyond the compactness
already produced by the evolved modified initial power spectrum.

Second, we repeat the calculation using the FCT reference profile, obtaining
\(M_{\rm sunk}^{\rm FCTinit}\) and \(r_{\rm eff}^{\rm FCTinit}\).  For this
reference profile we use the \(V_{\max}\)--\(R_{\max}\)-based infall
concentration, rather than the SO-based concentration, because the latter
develops a pathological high-concentration tail in stripped FCT subhaloes.
We then define
\[
\mu_{\rm FCT}
=
\frac{M_{\rm sunk}^{\rm FCTinit}}
     {M_{\rm NFW,FCT}(<r_{\rm eff}^{\rm FCTinit})}.
\]
This ratio measures the importance of the sunk component relative to the
already compact FCT reference profile.

Both \(\mu_{\rm CDMref}\) and \(\mu_{\rm FCT}\) are enclosed-mass
perturbation ratios evaluated at their corresponding \(r_{\rm eff}\); they
are not fractions of the PBH population that sinks.

\begin{figure}
    \centering
    \hspace{-.6cm}\includegraphics[width=0.93\linewidth]{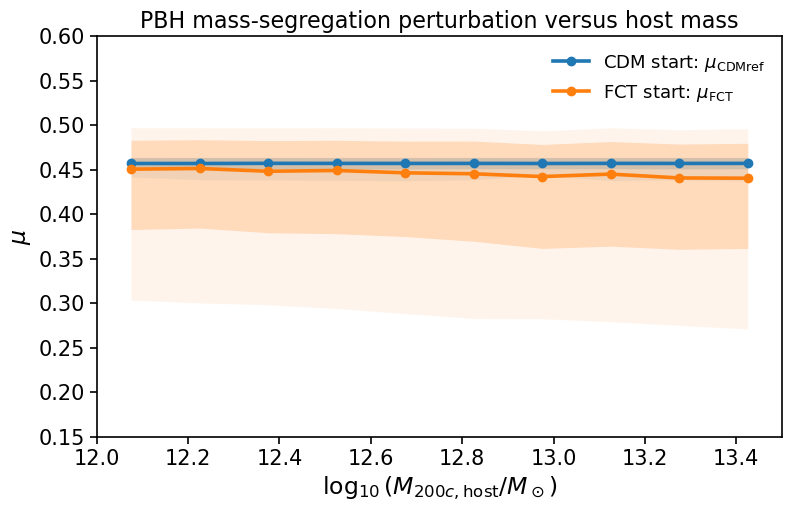}
    \includegraphics[width=0.95\linewidth]{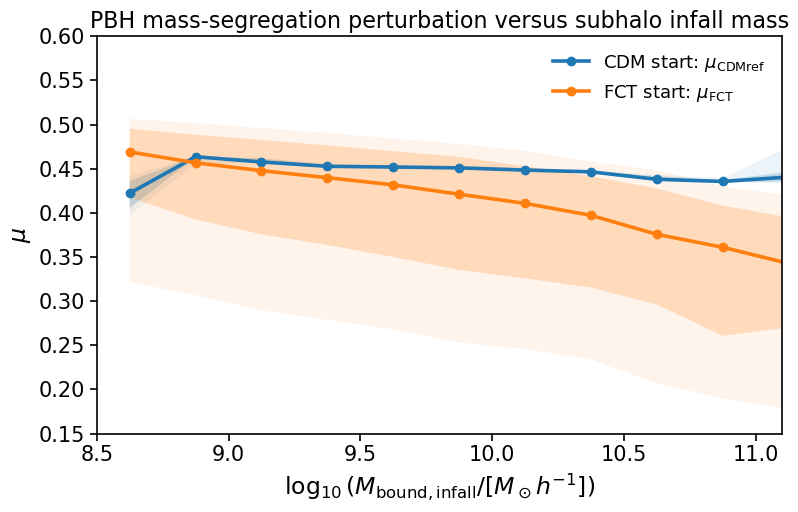}
\caption{
Median PBH mass-segregation perturbation as a function of host mass
(top) and subhalo infall bound mass (bottom).  Blue curves show the 
CDM-reference calculation, \(\mu_{\rm CDMref}\), and orange curves show the
FCT-reference calculation, \(\mu_{\rm FCT}\).  Points show medians and shaded
regions show the 16--84 and 5--95 percentile ranges.
}
    \label{fig:pbh_mu_trends}
\end{figure}
Figure~\ref{fig:pbh_mu_trends} shows little dependence on host mass, with
typical values \(\mu\simeq0.45\). Over the subhalo-mass range shown, the
CDM-reference perturbation is also nearly independent of
\(M_{\rm bound,infall}\), remaining close to
\(\mu_{\rm CDMref}\simeq0.4\)--0.5. By contrast,
\(\mu_{\rm FCT}\) decreases gradually with increasing infall mass, from
approximately \(0.45\)--0.5 at the low-mass end to approximately \(0.35\)
near \(M_{\rm bound,infall}\sim10^{11}\,M_\odot h^{-1}\). This decline occurs
because the sunk PBH component is compared with an increasingly compact FCT
reference profile. PBH segregation can therefore provide a substantial
central perturbation relative to either reference profile, but does not
generically dominate the already compact FCT inner mass.

\begin{figure}
    \centering
    \includegraphics[width=0.95\linewidth]{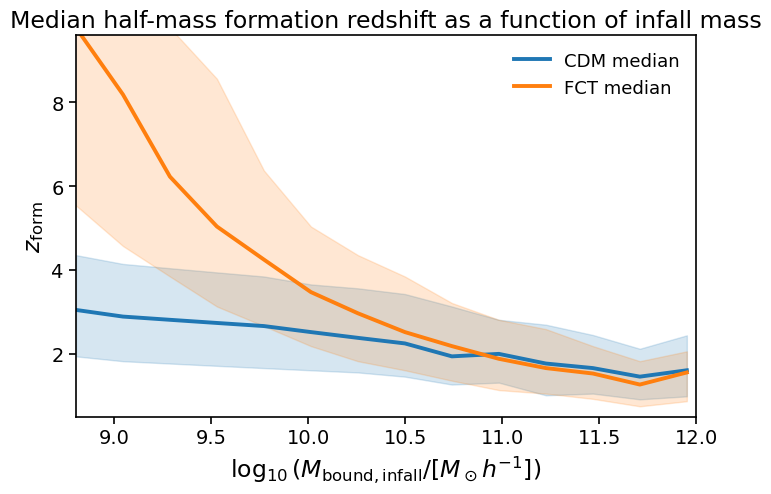}
   \caption{
Half-mass formation redshift as a function of subhalo infall bound mass.
Blue and orange curves show the median CDM and FCT relations, respectively,
and shaded regions show the 16--84 percentile intervals.
}
    \label{fig:zform}
\end{figure}

 Figure~\ref{fig:zform} shows that the enhanced small-scale fluctuations in
the FCT model substantially advance the assembly of low-mass subhaloes.
At \(M_{\rm bound,infall}\lesssim10^{10}\,M_\odot h^{-1}\), the median FCT
formation redshift is considerably higher than the corresponding CDM value,
whereas the two formation histories converge toward higher masses.  This
earlier assembly provides a natural contribution to the enhanced
concentration of FCT subhaloes: haloes that establish their inner structure
at higher redshift inherit a higher characteristic density.

 Analytic models and numerical simulations connect halo concentration to the
epoch at which the inner halo assembles, with simple prescriptions predicting
an approximately linear growth of concentration with the ratio of the
observation and formation scale factors
\citep{Bullock2001,Wechsler2002}.  Motivated by these results, we adopt the
first-order scaling
\[
c\propto 1+z_{\rm form},
\]
and define
\[
c_{\rm form}
=
c_{\rm CDM}
\left[
\frac{1+z_{\rm form,FCT}(M_{\rm bound,infall})}
     {1+z_{\rm form,CDM}(M_{\rm bound,infall})}
\right]^\eta,
\qquad \eta=1.
\]
This is a first-order approximation: the detailed relation between
concentration and assembly epoch depends on the formation-time definition and
the full mass-accretion history
\citep{Ludlow2014,Correa_2015}. In particular, our half-mass formation
redshift does not coincide exactly with the characteristic assembly epochs
used in those concentration models.

\begin{figure}
    \centering
    \includegraphics[width=0.95\linewidth]{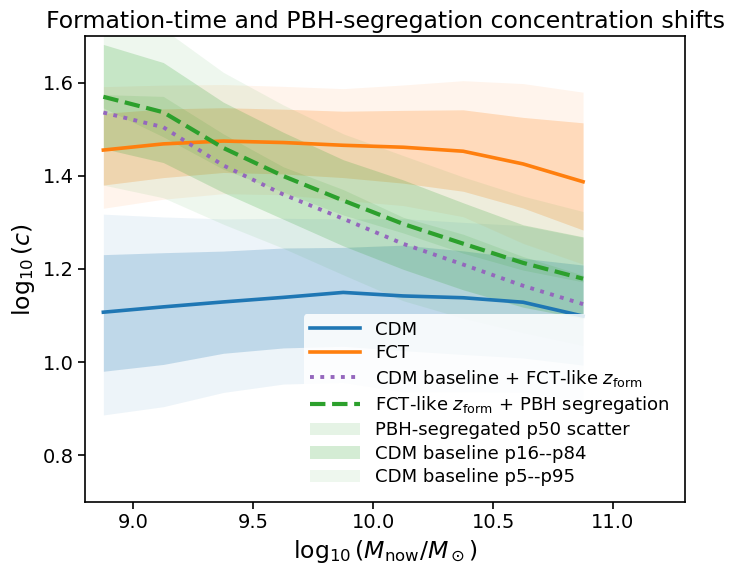}
    \caption{
    Present-day effective concentration-like proxy as a function of current
    bound-subhalo mass.  Blue and orange curves show the median CDM and FCT
    structural proxies, respectively.  The purple dotted curve shows the
    modified CDM baseline after imposing the earlier median FCT
    formation redshift at fixed \(M_{\rm bound,infall}\), assuming
    \(c\propto1+z_{\rm form}\).  The green dashed curve then applies the
    mass-conserving PBH-segregation calculation to this formation-time-shifted
    profile.  The green shaded regions show the propagated scatter and CDM
    baseline percentile ranges.  We define
    \(c_{\rm now,eff}^{\rm SOAP}=R_{200}(M_{\rm now})/r_{s,{\rm now}}\).
    Because the objects are stripped present-day satellites, this quantity is
    an NFW-equivalent compactness proxy and not a conventional
    spherical-overdensity concentration.
    }
    \label{fig:present_concentration}
\end{figure}

Figure~\ref{fig:present_concentration} shows that FCT subhaloes have
systematically larger present-day compactness proxies than CDM at fixed
\(M_{\rm now}\). The formation-time rescaling accounts for much of the
low-mass offset, where the CDM and FCT assembly histories differ most
strongly, but has little effect at high mass.

Second, we apply the PBH mass-segregation calculation to the
formation-time-shifted profile.  For each FCT subhalo, we solve for an
equivalent concentration \(c_{\rm form+seg}\) satisfying
\[
\begin{aligned}
&M_{\rm NFW}\!\left(
 <r_{\rm eff}^{\rm CDMref}\mid c_{\rm form+seg}
 \right)\\
&\quad =
\left(1-f_{\rm sunk}^{\rm CDMref}\right)
M_{\rm NFW}\!\left(
 <r_{\rm eff}^{\rm CDMref}\mid c_{\rm form}
 \right)
+M_{\rm sunk}^{\rm CDMref},
\end{aligned}
\]
where
\[
f_{\rm sunk}^{\rm CDMref}
=
\frac{M_{\rm sunk}^{\rm CDMref}}{M_{\rm now}}.
\]
The green dashed curve therefore includes the effects sequentially: the CDM
reference profile is first made more concentrated according to the earlier
FCT formation epoch, and the predicted segregated PBH mass is then added to
that modified profile.

 Earlier formation accounts for a substantial part of the CDM--FCT
concentration difference at low mass, while PBH segregation produces a
further increase.  The combined estimate nevertheless remains below the
median FCT compactness over most of the plotted range.  This indicates that
the measured FCT structural offset cannot be attributed solely to either the
simple formation-time rescaling or the unresolved segregated PBH component.
The remaining difference may reflect the full nonlinear response to the
modified initial power spectrum, departures from the assumed
\(c\propto1+z_{\rm form}\) mapping, and the subsequent tidal evolution of the
subhaloes.

\section{Abundance-corrected stellar-to-halo mass relation}
\label{sec:modified_shmr}

Subhalo abundance matching (SHAM) connects galaxies to dark-matter haloes by
assuming a monotonic relation between a galaxy property, such as luminosity
or stellar mass, and a halo or subhalo property.  Despite its simplicity,
SHAM reproduces galaxy clustering remarkably well when suitable halo proxies
are used \citep[e.g.][]{ValeOstriker2004,Conroy2006,Reddick2013}.  For
satellites, present-day bound mass is usually a poor proxy for stellar mass
because it is strongly affected by tidal stripping after accretion.
Quantities such as \(M_{\rm peak}\) or \(v_{\rm peak}\) are therefore
commonly preferred in object-by-object SHAM models, since they better trace
the depth of the subhalo potential before stripping
\citep{Reddick2013,Campbell2018}.

In this work, however, our goal is not to construct a calibrated
object-by-object SHAM catalogue.  Instead, we ask how the stellar-to-halo
mass relation inferred in CDM would be modified if the cumulative abundance
of subhaloes were changed by PBH/FCT physics.  We therefore use
the progenitor halo mass at last isolation,
\(M_{\rm vir,sub}^{\rm infall}\), as the mass variable for the
stellar--halo mass assignment.  This quantity is obtained from the SOAP
spherical-overdensity mass \(M_{200{\rm c},\rm sub}^{\rm infall}\), converted
to the virial mass convention used by the \citet{Behroozi13} relation.  The
HBT bound mass at infall, \(M_{\rm bound,infall}\), is used only to display
the raw bound-substructure abundance in this section; the abundance-corrected
stellar-mass assignment itself is constructed in
\(M_{\rm vir,sub}^{\rm infall}\). This choice allows us to compare the CDM
and FCT cumulative subhalo abundances at fixed progenitor virial
mass and to propagate this difference directly into an abundance-based
correction to the \citet{Behroozi13} stellar-to-halo mass relation.

Using \(V_{\max,{\rm peak}}\) would be appropriate for assigning stellar
masses to individual subhaloes in a standard SHAM analysis, but it is less
suited to the abundance mapping used here.  It would require choosing a
velocity-based mapping between \(V_{\max,{\rm peak}}\) and stellar mass, and
would mix the abundance effect with changes in concentration
and internal structure.  These effects are themselves modified in
the FCT model and are treated separately through the enclosed-mass profile
comparison below.  We therefore use \(M_{\rm vir,sub}^{\rm infall}\)
to define the abundance-corrected stellar--halo mass relation, while using
present-day structural quantities such as \(V_{\max,{\rm now}}\) and
\(R_{\max,{\rm now}}\) only when constructing the present-day
bound-remnant mass profiles for the dynamical-mass comparison.

The analysis is performed on host haloes with masses above
\(\sim10^{12}\,M_\odot\), where host masses are SOAP
\(M_{200{\rm c},{\rm host}}\) values in physical solar masses, in the CDM and
FCT simulations.  We also consider narrower or higher-mass host selections as
a check on host-mass dependence.  Because the simulation volumes are small,
these samples are not large enough to provide a statistically representative
halo or subhalo mass function over the full mass range of interest.  We
therefore do not attempt a full-volume abundance-matching calibration, and
use the CDM stellar-mass assignment as a reference and estimate
the relative FCT correction from the cumulative abundance of subhaloes as a
function of \(M_{\rm vir,sub}^{\rm infall}\).

We first compute the raw cumulative infall-bound subhalo abundances per host,
\[
\left\langle N_{\rm sub}(>M_{\rm bound,infall})\right\rangle_{\rm host},
\]
for CDM and FCT host samples selected by host mass. This quantity
uses the HBT bound mass at infall, and is used to show
how the resolved bound-substructure population changes between CDM and FCT.
For each host selection we compute
\[
\left\langle
N_{\rm CDM}(>M_{\rm bound,infall})
\right\rangle_{\rm host}
\quad \text{and} \quad
\left\langle
N_{\rm FCT}(>M_{\rm bound,infall})
\right\rangle_{\rm host}.
\]
These curves are used as a direct comparison of the relative CDM and
FCT bound-substructure populations. For the abundance-corrected
stellar-mass assignment itself, we use the analogous cumulative abundance as
a function of \(M_{\rm vir,sub}^{\rm infall}\), so that the abundance mapping
and the Behroozi relation use the same halo-mass convention.  The CDM and FCT
cumulative abundances differ in shape, not only in normalization, because the
FCT/PBH component changes the small-scale initial power spectrum and the resulting
subhalo population.

\begin{figure}
    \centering
    \includegraphics[width=0.95\linewidth]{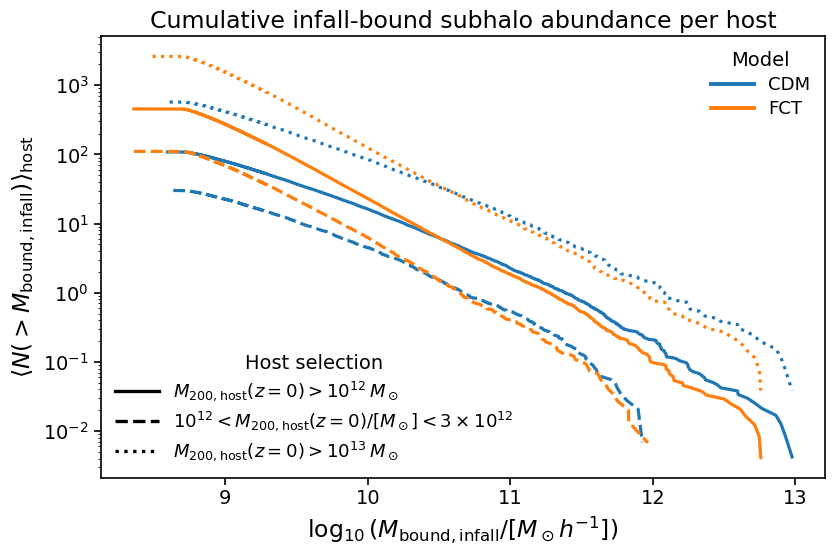}
\caption{
Raw host-averaged cumulative infall-bound subhalo abundance for CDM and
FCT.  Curves show
\(\langle N_{\rm sub}(>M_{\rm bound,infall})\rangle_{\rm host}\), where
\(M_{\rm bound,infall}\) is the HBT bound mass at infall.  Line styles show
different present-day SOAP host-mass selections:
\(M_{200{\rm c},{\rm host}}>10^{12}\,M_\odot\) (solid),
\(10^{12}<M_{200{\rm c},{\rm host}}/M_\odot<10^{12.5}\) (dashed), and
\(M_{200{\rm c},{\rm host}}>10^{13}\,M_\odot\) (dotted).
}
    \label{fig:cmf_infall_bound}
\end{figure}

Figure~\ref{fig:cmf_infall_bound} shows the cumulative abundance of
subhaloes as a function of their bound mass at infall for three present-day
host-mass selections.  The overall normalization increases with host mass, as
expected, but the relative CDM--FCT difference is broadly similar across the
three selections.  In all cases, FCT enhances the abundance of low-mass
infall subhaloes relative to CDM, while the difference becomes smaller toward
the high-mass end.  We therefore treat the FCT modification to the infall
subhalo abundance as approximately independent of present-day host halo mass
over the range used here.

The massive-subhalo tail is sparsely sampled and shows small crossings
between the CDM and FCT curves, so we do not use it to define a separate
host-mass-dependent correction.  What we do is use the full cumulative subhalo
distribution in \(M_{\rm vir,sub}^{\rm infall}\) to define the abundance-based
stellar-mass assignment below.

We take the \citet{Behroozi13} stellar-to-halo mass relation as the baseline
CDM mapping by evaluating it at the infall redshift of each subhalo.  For a
CDM subhalo we assign
\[
M_{\star,i}^{\rm CDM}
=
M_\star^{\rm Behroozi}
\left(M_{{\rm vir,sub},i}^{\rm infall},z_{{\rm infall},i}\right).
\]
Here \(M_{{\rm vir,sub},i}^{\rm infall}\) is obtained from the SOAP
\(M_{200{\rm c},\rm sub}^{\rm infall}\) value at the HBT-HERONS
last-isolation snapshot, converted to the virial convention used by
\citet{Behroozi13}. This choice is closer to the regime in which the
\citet{Behroozi13} relation is calibrated for central galaxies, because it
associates the satellite stellar mass with the halo mass near accretion,
before the subhalo loses a large fraction of its dark matter through tidal
stripping.  After infall, star formation in satellites is often reduced or
quenched, so the stellar mass is expected to evolve much less than the bound
dark-matter mass.  In addition, stars are more centrally concentrated and more
tightly bound than the dark matter, so tidal stripping removes dark matter
preferentially before strongly affecting the stellar component
\citep{Smith2016,Errani2022}.  The progenitor virial mass and infall
redshift therefore provide a physically motivated proxy for the halo
properties that set the satellite stellar mass.  This gives a consistent CDM
baseline for defining the relative FCT abundance correction.

\begin{figure}
    \centering
    \includegraphics[width=0.95\linewidth]{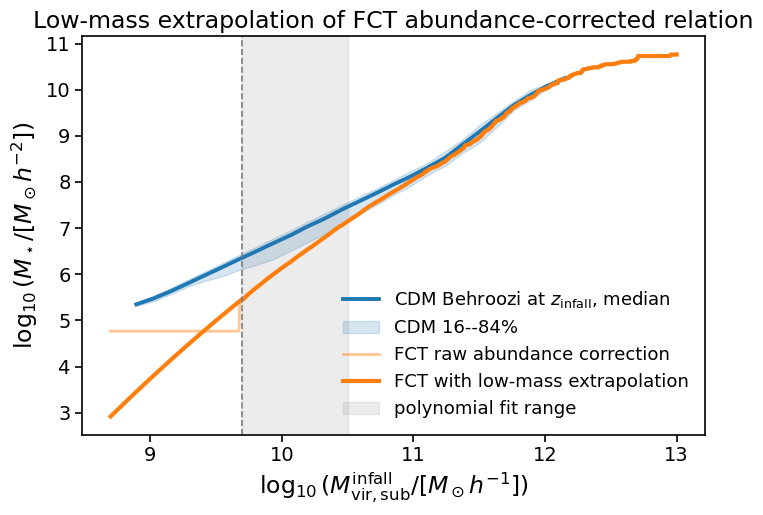}
\caption{
Abundance-based correction to the CDM stellar-to-halo mass relation using
the progenitor virial mass at infall,
\(M_{\rm vir,sub}^{\rm infall}\).  The blue curve and band show the CDM
baseline from the redshift-dependent \citet{Behroozi13} relation.  The
light-orange curve shows the direct FCT abundance-matched mapping, and the
dark-orange curve shows the smooth low-mass extension used for the Milky Way
satellite comparison.
}
    \label{fig:modified_behroozi}
\end{figure}

The CDM cumulative stellar-mass abundance is then obtained by applying this
redshift-dependent Behroozi mapping to the CDM subhalo catalogue in the
selected host sample and counting the number of subhaloes with
\(M_{\star,i}^{\rm CDM}>M_\star\).  At fixed
\(M_{\rm vir,sub}^{\rm infall}\), the CDM assignment is not strictly
single-valued, because different subhaloes have different infall redshifts.
The blue band shown in Figure~\ref{fig:modified_behroozi} therefore reflects
the range of CDM stellar masses induced by the distribution of
\(z_{\rm infall}\).

For the FCT catalogue we define an abundance-corrected stellar mass by
matching each FCT subhalo to the same cumulative abundance in the CDM
baseline,
\[
N_{\star}^{\rm CDM}(>M_{\star}^{\rm FCT})
=
N_{\rm FCT}(>M_{\rm vir,sub}^{\rm infall}) .
\]
Equivalently, the FCT stellar mass assigned to
\(M_{\rm vir,sub}^{\rm infall}\) is the stellar mass in the CDM
baseline that has the same cumulative abundance, or, in practice, the same
cumulative number per host in the selected sample.  Since the CDM baseline is
built from the redshift-dependent Behroozi assignment, the resulting FCT
abundance-corrected mapping also inherits this \(z_{\rm infall}\)-dependent
reference statistically, while preserving the change in cumulative subhalo
abundance produced by the FCT model.

This prescription isolates the abundance effect of the FCT subhalo mass
function.  If FCT has more subhaloes than CDM above a given
\(M_{\rm vir,sub}^{\rm infall}\), the same cumulative stellar-mass
abundance implies a lower assigned \(M_\star\) at fixed infall mass.
Conversely, in mass ranges where FCT has fewer subhaloes than CDM, the
abundance-matched FCT relation shifts to higher stellar mass.  

Figure~\ref{fig:modified_behroozi} shows the resulting abundance-corrected
mapping.  The light-orange curve is obtained directly by matching the raw FCT
cumulative subhalo abundance to the redshift-dependent CDM stellar-mass
baseline.  Its flattening or irregular behaviour at the lowest masses reflects
the finite lower limit of the available CDM catalogue and the discrete nature
of the cumulative counts.  For the Milky Way satellite comparison we also use
a smooth low-mass continuation, obtained by fitting the abundance-corrected
relation over the shaded interval and extrapolating to lower
\(M_{\rm vir,sub}^{\rm infall}\).  This continuation provides a
smooth extension of the same abundance-based mapping into the ultra-faint
satellite regime.

The CDM--FCT abundance-matching offset becomes most relevant below
\(M_{\rm vir,sub}^{\rm infall}\sim10^{10.5}\,M_\odot\), outside the directly
calibrated range of the Behroozi relation. This is also the regime where
ultra-faint satellites are usually modelled with stochastic occupation,
scatter, and survey selection effects
\citep[e.g.][]{Simon2019,Nadler2020,SantosSantos2022}.
Several physical mechanisms could plausibly produce a reduced stellar-mass
conversion efficiency in these low-mass FCT haloes. In particular, their
earlier formation allows star formation to begin in lower-mass progenitors
at higher redshift, where feedback from the first generations of stars,
including Pop~III stars, can heat or expel gas from shallow potential wells.
The enhanced abundance and earlier formation of low-mass objects could also
modify the timing and thermal history of reionization, after which
photoheating suppresses gas retention and reaccretion in shallow potential
wells. These effects could reduce the integrated star-formation efficiency
of the progenitors of low-mass FCT haloes relative to later-forming systems
in \(\Lambda\)CDM. Additional heating associated with accretion onto massive
PBHs could provide a further suppression channel. These baryonic processes
and the reionization history are not modelled by the dark-matter-only
simulations considered here. The abundance-corrected relation should
therefore be regarded as an effective mapping that quantifies the change
required by the modified subhalo abundance, rather than as a physical
prediction for the FCT stellar-to-halo mass relation.

The main result of the CDM--FCT offset is therefore not a simple uniform shift in
the stellar-to-halo mass relation.  Because the FCT and CDM cumulative
subhalo abundances have different shapes, the abundance-corrected FCT mapping
is mass dependent: at fixed \(M_{\rm vir,sub}^{\rm infall}\), it
moves below the CDM baseline where FCT is more abundant than CDM, and above it
where FCT is less abundant.  This mass dependence is the key feature
propagated into the satellite stellar-mass-function comparison below.

\section{Comparison with Milky Way satellites}
\label{sec:mw_satellites}

We now compare the CDM and FCT predictions with two Milky Way satellite
diagnostics.  We analyse the predicted cumulative satellite stellar-mass function, for which we use the
Milky Way satellite census compiled by \citet{DrlicaWagner2020}, retaining
their class 3 and class 4 systems.  This gives \(N_{\rm obs}=57\) satellites
and is used without any completeness correction, so the observed
stellar-mass function provides a lower limit to the intrinsic
Milky Way satellite abundance.  And we also perform an internal dynamical comparison, for which we use
the Local Volume Database (LVDB) compilation and retain the subset of
satellites with measured half-light radii and dynamical masses, giving
\(N_{\rm dyn}=38\) systems.  The cumulative stellar-mass function tests the
combined effect of the subhalo abundance and the assumed stellar-to-halo mass
mapping.  The half-light-radius--dynamical-mass relation instead probes the
internal density structure of the subhaloes.

\subsection{Stellar mass functions}

We first compare the cumulative stellar-mass function of the
\citet{DrlicaWagner2020} class 3 and class 4 sample with the simulated
satellite populations. For the simulated catalogues, we select all hosts in a Milky-Way-like interval
using the SOAP present-day host mass,
\[
11.8 < \log_{10}(M_{200{\rm c},{\rm host}}/M_\odot) < 12.2 .
\]

\begin{figure}
    \centering
    \includegraphics[width=0.95\linewidth]{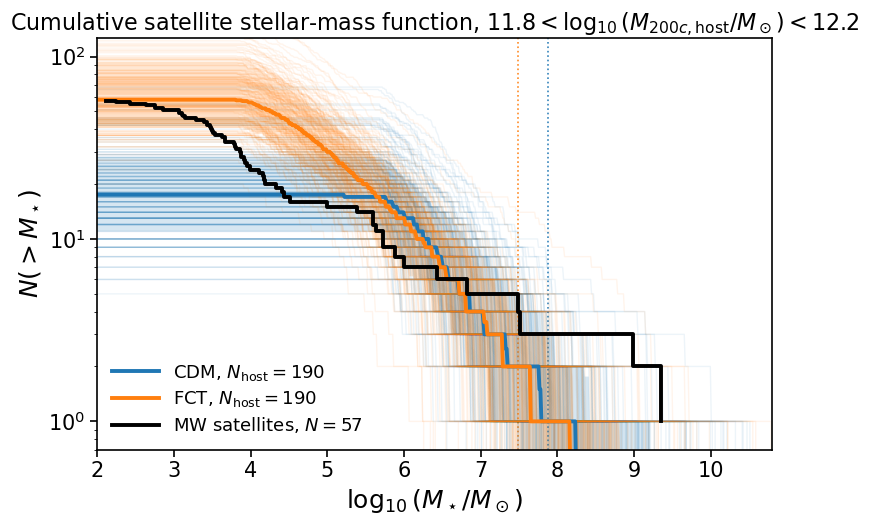}
\caption{
Cumulative satellite stellar-mass function for Milky-Way-mass hosts.
Thin blue and orange curves show individual CDM and FCT hosts in the
interval
\(11.8 < \log_{10}(M_{200{\rm c},{\rm host}}/M_\odot) < 12.2\). Thick blue and orange curves show the median relation across hosts, and
shaded bands indicate the 16--84 percentile host-to-host range.  The black curve shows the observed Milky Way satellite census from
\citet{DrlicaWagner2020}, using their class 3 and class 4 systems
(\(N_{\rm obs}=57\)), without correcting for incompleteness.  Stellar masses are assigned using
the original \citet{Behroozi13} relation for CDM and the abundance-corrected
FCT relation from Sect.~\ref{sec:modified_shmr}.  The vertical dotted lines
mark the stellar masses corresponding to
\(M_{\rm bound,infall}=10^{10.5}\,M_\odot h^{-1}\) in the CDM and FCT
mappings, respectively; below these values the plotted relations rely on
low-mass extrapolation.  
}
    \label{fig:sat_smf_mw}
\end{figure}
 
Figure~\ref{fig:sat_smf_mw} compares the observed Milky Way satellite
stellar-mass census with the simulated Milky-Way-mass host sample. The thin
curves show the substantial host-to-host variation in both cosmologies. CDM
host realizations overlap the observed abundance of the most massive Milky
Way satellites somewhat more frequently. The FCT median predicts a similar
number of satellites over much of the plotted mass range, although fewer
individual FCT systems reproduce the high-mass end. Inspection of individual
hosts nevertheless shows that matching the most massive Milky Way satellites
does not necessarily imply an overall excess of satellites: some hosts contain
a few massive satellites but their cumulative abundance then approaches the
sample median at lower stellar masses, producing a shape qualitatively similar
to that of the Milky Way. Thus the massive Milky Way satellites can be
reproduced in both cosmologies, although only by a relatively small fraction
of Milky-Way-mass hosts.

Over the stellar-mass range where the mapping is better constrained, the CDM
and FCT median satellite stellar-mass functions remain broadly similar in
shape, despite the larger FCT normalization.  The largest FCT excess appears
toward the lowest stellar masses.  Thus the strong differences in the
underlying subhalo abundance do not translate into a clean constraint from
the luminous satellite stellar-mass function alone.  The abundance-corrected
mapping absorbs part of the difference by assigning stellar masses according
to cumulative abundance, rather than by applying a fixed stellar-to-halo mass
relation at fixed halo mass.

The largest CDM--FCT differences occur where both the CDM and FCT mappings
rely on low-mass extrapolation and where the observed
\citet{DrlicaWagner2020} census is affected by incompleteness from sky
coverage, surface-brightness limits, and distance-dependent selection
effects.  We therefore do not interpret the low-mass divergence as a direct
prediction for the number of observable ultra-faint satellites, 
but this is the stellar-mass range where the modified FCT subhalo abundance would have
the largest impact if low-mass haloes are able to form and retain stars
efficiently.

The structural comparison discussed in Sect.~\ref{sec:halo_catalogues}
provides the complementary information: FCT subhaloes are more compact in the
resolved SOAP circular-velocity reconstruction.  The stellar-mass function
therefore tests the abundance side of the model, while the enclosed-mass
comparison below tests whether the corresponding candidate satellite hosts
are dynamically compatible with the measured masses of Milky Way satellites.

\subsection{Internal dynamical masses}

We then compare the internal dynamical masses of Milky Way satellites with
the enclosed-mass profiles predicted for simulated subhaloes.  This
comparison requires measured line-of-sight velocity dispersions and therefore
cannot be performed for the full \citet{DrlicaWagner2020} satellite census.
We use the Local Volume Database (LVDB) compilation of nearby dwarf galaxies
and star clusters \citep{Pace2024LVDB}, selecting Milky Way satellites that
are classified as confirmed galaxies and have measured projected half-light
radii and line-of-sight velocity dispersions.  We exclude systems with
velocity-dispersion upper limits only, and apply mild size and luminosity cuts,
\(R_{\rm e}>10\,{\rm pc}\) and \(M_V<-1\), to remove compact star-cluster
contaminants while retaining the diffuse ultra-faint dwarf population.
After these cuts, the dynamical comparison sample contains
\(N_{\rm dyn}=38\) confirmed Milky Way satellites.

The observed half-light radii in this sample are generally  smaller than the
simulation force-softening length, which is \(1.8\) comoving kpc.  We therefore
do not measure \(M(<R_{\rm e})\) directly from simulation particles.  Instead,
for each simulated subhalo we infer an approximate present-day bound-remnant
profile from \(V_{\max,{\rm now}}\) and \(R_{\max,{\rm now}}\), and evaluate
this profile at the assigned stellar half-light radius. The comparison is therefore a Too-Big-To-Fail-like (TBTF) internal structure test
\citep{BoylanKolchin2011TBTF,BoylanKolchin2012MW} based on the resolved
\(V_{\max}\) and \(R_{\max}\) of the subhaloes.

We estimate the dynamical mass within the deprojected half-light radius using
the estimator of \citet{Wolf2010},
\[
M_{1/2}\simeq \frac{4}{G}\sigma_{\rm los}^2R_{\rm e}
\simeq
930\,
\left(\frac{\sigma_{\rm los}}{{\rm km\,s^{-1}}}\right)^2
\left(\frac{R_{\rm e}}{{\rm pc}}\right)
M_\odot ,
\]
where \(R_{\rm e}\) is the projected half-light radius and
\(\sigma_{\rm los}\) is the line-of-sight velocity dispersion.

\begin{figure}
    \centering
    \includegraphics[width=0.98\linewidth]{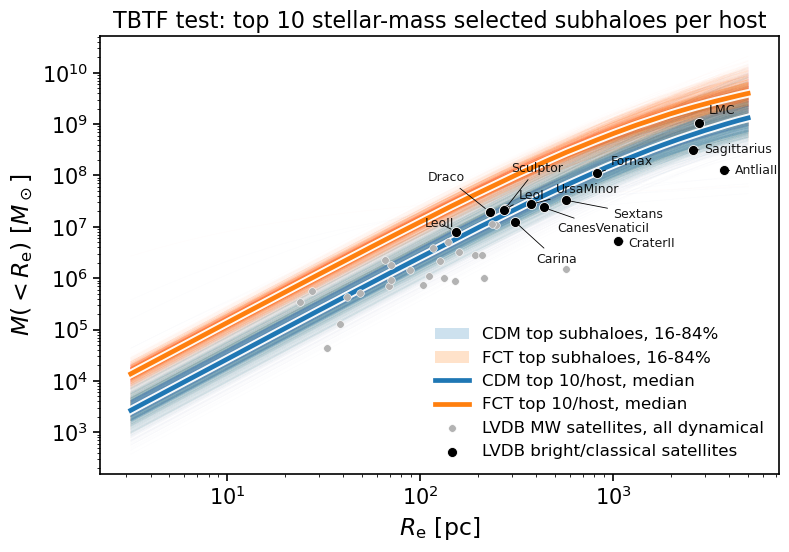}
    \includegraphics[width=1.\linewidth]{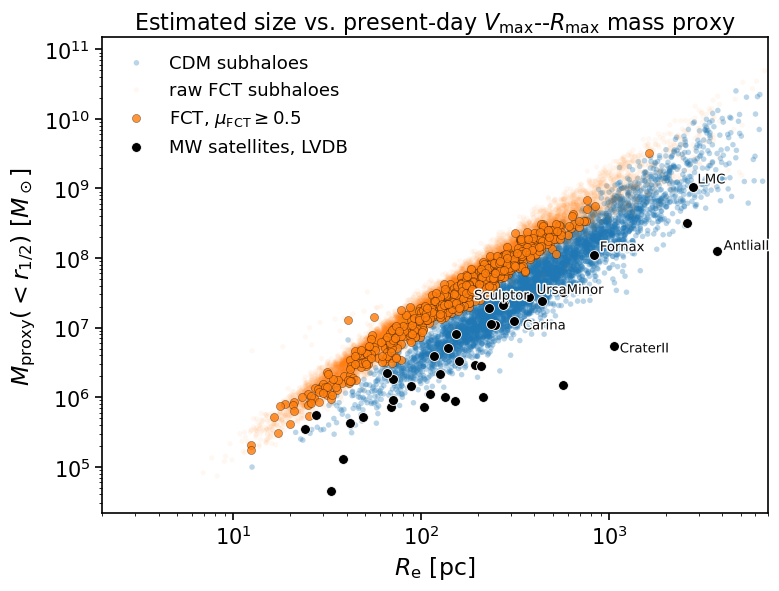}
\caption{
Comparison between observed Milky Way satellite dynamical masses and
simulated subhalo enclosed masses.  Top panel: LVDB satellites are shown as
points, and blue/orange curves show NFW enclosed-mass profiles for the top
ten stellar-mass-selected CDM and FCT subhaloes per Milky-Way-mass host,
reconstructed from present-day SOAP \(V_{\max}\) and \(R_{\max}\).  Thick
curves show the median profile, shaded bands indicate the 16--84 percentile
range, and faint curves show individual selected subhaloes.  Bottom panel:
one-point-per-subhalo comparison using assigned stellar sizes and the same
present-day \(V_{\max}\)--\(R_{\max}\) mass proxy.  Highlighted orange points
mark FCT subhaloes with \(\mu_{\rm FCT}>0.5\).
}
    \label{fig:sizes}
\end{figure}

For the simulations, we select Milky-Way-mass hosts in the same host-mass
interval as above and rank their subhaloes by the assigned stellar mass.  For
CDM this stellar mass is obtained from the redshift-dependent
\citet{Behroozi13} assignment, while for FCT it is obtained from the
abundance-corrected mapping of Sect.~\ref{sec:modified_shmr}.  In the top
panel of Fig.~\ref{fig:sizes} we show only the top ten subhaloes per host.
This gives a TBTF-style comparison between the most likely hosts of the
brightest satellites and the observed dynamical masses of luminous Milky Way
satellites \citep{BoylanKolchin2011TBTF,BoylanKolchin2012MW}.  The
enclosed-mass profiles are reconstructed from the present-day SOAP
\(V_{\max}\) and \(R_{\max}\), assuming an NFW form.

The simulated profiles in Fig.~\ref{fig:sizes} are NFW reconstructions
anchored to the present-day \(V_{\max}\) and \(R_{\max}\) of each bound
subhalo.  They therefore compare the effective compactness of CDM and FCT
satellite remnants on resolved circular-velocity scales.  In this diagnostic the raw FCT
population is more compact than CDM, producing larger enclosed masses at
fixed radius.

The subgrid PBH mass-segregation calculation is used here only to identify
where the unresolved PBH population would make an additional inner-mass
contribution.  In the FCT model used here, PBHs in the high-mass tail of the
extended mass function can sink through dynamical friction against the
lower-mass PBH background and the collective halo potential.  The diagnostics
\(\mu_{\rm CDMref}\) and \(\mu_{\rm FCT}\) compare the predicted sunk PBH
mass with a  CDM-reference profile and with the FCT reference profile,
respectively.  In each case the sunk mass and effective sinking radius are
computed self-consistently for the corresponding starting profile.  In Fig.~\ref{fig:sizes}
we use \(\mu_{\rm FCT}\) to highlight FCT subhaloes for which this sunk
component is an order-unity perturbation relative to the already compact FCT
reference profile.

The top panel of Fig.~\ref{fig:sizes} shows that the selected CDM subhaloes
are broadly comparable to the bright/classical LVDB locus, although the
densest systems still show a mild TBTF-like excess
\citep{BoylanKolchin2011TBTF,BoylanKolchin2012MW}.  The selected FCT
subhaloes are systematically shifted toward larger enclosed masses at fixed
radius, producing a denser median and 16--84 percentile range than CDM.  This
shift moves the FCT candidates above most of the observed bright/classical
satellites, including the LMC-scale point, rather than improving the match to
the observed locus.  Treating all compact FCT subhaloes as candidate luminous
satellite hosts would therefore strengthen the TBTF-like tension.

However, the simulated curves are dark-matter-only NFW reconstructions based
on present-day \(V_{\max}\) and \(R_{\max}\), without baryonic disruption,
stellar response, or a probabilistic galaxy--subhalo occupation model.
Accordingly, this comparison should be interpreted as a structural-density
diagnostic of the candidate subhalo population, not as a direct prediction
for the observed Milky Way satellite mass--size relation.

The bottom panel of Fig.~\ref{fig:sizes} shows a complementary
one-point-per-subhalo comparison.  To do so we assign each subhalo an
effective radius \(R_{200}(M_{\rm bound,infall})\), computed from the
infall-bound mass using the \(z=0\) critical density, and a
concentration-dependent Jiang-like scaling,
\[
r_{1/2,\star}
=
0.020
\left(\frac{c_{\rm infall}}{10}\right)^{-0.7}
R_{200}(M_{\rm bound,infall}) ,
\]
motivated by empirical and simulated galaxy-size--halo-radius relations
\citep{Kravtsov2013,Jiang2019}.  We then convert to a projected half-light
radius using
\[
R_{\rm e}=0.75\,r_{1/2,\star},
\]
consistent with the deprojection convention used in the dynamical-mass
estimator of \citet{Wolf2010}.  The enclosed mass is evaluated at the
assigned three-dimensional half-light radius, \(r_{1/2,\star}\), using the
same present-day bound-remnant \(V_{\max,{\rm now}}\) and
\(R_{\max,{\rm now}}\) NFW reconstruction.

The full FCT population is shifted toward larger enclosed masses than CDM at
fixed assigned size, consistent with the direct profile comparison in the top
panel. We highlight the subset with the largest relative contribution from
sunk PBHs in the Milky-Way-mass host sample. Since \(\mu_{\rm FCT}\) reaches
values of order unity only marginally in this sample, we adopt
\[
\mu_{\rm FCT}
=
\frac{M_{\rm sunk}^{\rm FCTinit}}
     {M_{\rm NFW,FCT}(<r_{\rm eff}^{\rm FCTinit})}
>0.5
\]
as a practical high-\(\mu_{\rm FCT}\) threshold. These systems do not define
a separate sequence from the full FCT population. Instead, they occupy the
same broad FCT locus, with a preference for smaller assigned radii. This
indicates that the subhaloes most susceptible to additional PBH
mass-segregation effects are preferentially found in the compact,
low-\(R_{\rm e}\) part of the assigned-size population. Their subsequent
stellar response is not determined by the present calculation: a compact
central PBH component could deepen the central potential and promote
contraction, whereas dynamical friction and encounters with sinking massive
PBHs could transfer orbital energy to the stellar component and produce
heating or expansion.

The highlighted systems therefore do not define a distinct sequence that
brings the FCT population into better agreement with the Milky Way satellite
size--mass relation.  They only identify where PBH mass segregation is
expected to be most important within the already compact FCT population and where additional signatures associated with a central PBH population are most likely to arise.  The
bottom panel is model-dependent, because both the assigned stellar sizes and
the mapping between compact subhaloes and visible satellites depend on the
adopted size and occupation prescriptions.  

\begin{figure*}
    \sidecaption
    \includegraphics[width=12.9cm]{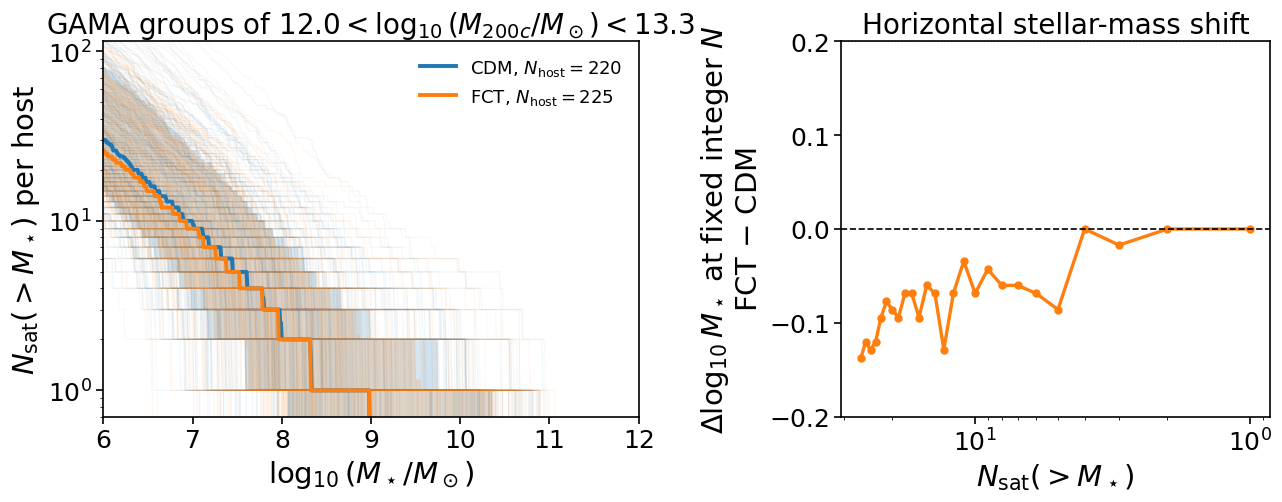}
    \caption{
    GAMA-like group-scale consistency check.  Left: cumulative satellite
    stellar mass functions for simulated hosts in the low-mass group interval
    \(12.0 < \log_{10}(M_{200{\rm c},{\rm host}}/M_\odot) < 13.3\).
    Thin curves show individual hosts, while thick curves show the median CDM
    and FCT predictions.  Right: horizontal stellar-mass shift between the FCT
    and CDM median curves at fixed cumulative abundance.  Positive values
    indicate that the FCT mapping assigns larger stellar masses at fixed
    satellite abundance.  
    }
    \label{fig:gama_like_groups}
\end{figure*}

\section{Beyond Milky Way satellites}
\label{sec:beyond_mw}

\subsection{GAMA group satellite stellar mass functions}
\label{sec:gama_groups}

As a higher-mass check on the abundance mapping, we examine simulated hosts in
the halo-mass range where GAMA group satellite stellar-mass functions have
been measured \citep{VazquezMata2020}.  Our simulated hosts overlap the
lowest GAMA halo-mass bin,
\(12.0 < \log_{10}(M_{200{\rm c},{\rm host}}/M_\odot) < 13.3\), which
corresponds to low-mass groups rather than Milky-Way analogues.

For this check, we apply the redshift-dependent CDM Behroozi relation and the
FCT abundance-corrected relation to the progenitor virial masses at last
isolation, using the same mass convention as in
Sect.~\ref{sec:modified_shmr}.  Figure~\ref{fig:gama_like_groups} shows the
resulting cumulative satellite stellar mass functions per host.  We do not
overplot the \citet{VazquezMata2020} fit for the lowest GAMA group bin,
because their measurement is calibrated mainly above
\(M_\star\sim10^9\,M_\odot\), while the simulated satellite counts in our
measurements are dominated by lower stellar masses.

The CDM and FCT median satellite stellar-mass functions are very similar over
the well-sampled range. The right-hand panel quantifies this comparison as a
horizontal stellar-mass shift at fixed integer cumulative satellite count.
For most of the overlap region, the offset remains within
\(-0.15\lesssim\Delta\log_{10}M_\star\lesssim0\), corresponding to at most a
\(\sim30\%\) reduction in stellar mass at fixed satellite abundance. The
offset approaches zero toward the smallest satellite counts, where the median
curves are determined by only a few massive satellites.

Thus, after applying the abundance-corrected FCT mapping, low-mass group
satellite stellar-mass functions do not require a substantial change in the
stellar-to-halo mass conversion relative to CDM. A cleaner test would likely
require satellite stellar-mass functions in lower-mass hosts, closer to or
below the Milky-Way-mass regime, where the relative change in the subhalo
population is expected to be more important. This is observationally
challenging, because constructing statistical samples of satellite systems
around hosts below \(\sim10^{12}\,M_\odot\) is difficult.

\subsection{Implications for compact dark substructure in massive hosts}
\label{sec:dark_substructure}

The subgrid PBH mass-segregation estimate also has implications beyond the
luminous satellite population.  In the PBH-FCT model considered here, with
\(M_\ast=100\,M_\odot h^{-1}\), \(f_{\rm PBH}=1\), and \(n_b=2.5\), the
simulation does not resolve individual PBHs.  It predicts the abundance and
\(V_{\max}\)--\(R_{\max}\)-based structure of FCT subhaloes; the subgrid
calculation identifies which of these subhaloes could also receive a sizeable
inner perturbation from the sinking high-mass tail of the unresolved PBH
population.

\begin{figure}
    \centering
    \includegraphics[width=0.95\linewidth]{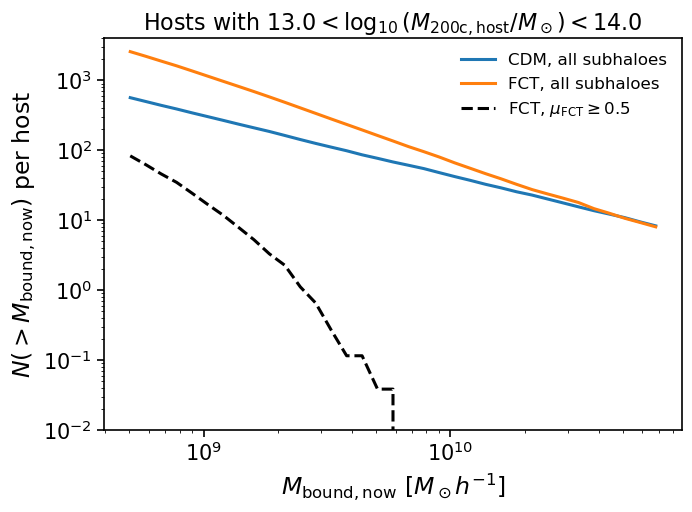}
\caption{
Host-averaged cumulative abundance of subhaloes in the lensing-motivated host
sample,
\(13.0<\log_{10}(M_{200{\rm c},{\rm host}}/M_\odot)<14.0\).
Blue and orange curves show the total CDM and FCT cumulative abundances as
functions of present-day bound subhalo mass,
\(M_{\rm bound,now}\). The dashed black curve shows the subset of FCT
subhaloes with a large PBH mass-segregation perturbation,
\(\mu_{\rm FCT}>0.5\), where
\(\mu_{\rm FCT}=M_{\rm sunk}^{\rm FCTinit}/
M_{\rm NFW,FCT}(<r_{\rm eff}^{\rm FCTinit})\).
}
    \label{fig:pdense_mass_function_highhosts}
\end{figure}

We examine this PBH-affected substructure population in a
lensing-motivated host sample,
\(13.0<\log_{10}(M_{200{\rm c},{\rm host}}/M_\odot)<14.0\).  This range is
closer to the group-scale environments of massive galaxy lenses.
Figure~\ref{fig:pdense_mass_function_highhosts} shows that the FCT model
produces an enhanced abundance of low-mass present-day bound subhaloes
relative to CDM in this host sample. The enhancement is strongest below
approximately \(10^{10}\,M_\odot h^{-1}\) and decreases toward higher masses,
where the two cumulative abundances approach one another. The FCT subhaloes
are also more compact than CDM in the \(V_{\max}\)--\(R_{\max}\)-based
structural proxy.

The dashed curve shows the subset of FCT subhaloes for which the 
FCT-reference PBH mass-segregation diagnostic satisfies
\[
\mu_{\rm FCT}
=
\frac{M_{\rm sunk}^{\rm FCTinit}}
     {M_{\rm NFW,FCT}(<r_{\rm eff}^{\rm FCTinit})}
>0.5 .
\]
This criterion does not require the sunk PBH component to dominate over the
FCT reference enclosed mass.  Instead, it selects systems in which the
unresolved PBH high-mass tail contributes an order-half inner perturbation
relative to the already compact FCT bound-remnant proxy. 
The \(\mu_{\rm FCT}>0.5\) population is a small subset of the total FCT
subhalo abundance and is restricted primarily to present-day bound masses
below a few \(10^9\,M_\odot h^{-1}\). These objects identify the low-mass tail
of the surviving FCT substructure population in which unresolved PBH mass
segregation could make the largest additional contribution to the inner mass
profile.

This possibility is most directly connected to gravitational probes of dark
substructure.  Strong-lensing constraints on low-mass structure are obtained
from massive galaxy-scale lenses and from haloes along the line of sight
\citep{dalal_direct_2002,Vegetti_2009,VegettiReview2024,Despali_2025}.
These probes are sensitive to the projected mass distribution near the lensed
images, and therefore to both abundance and compactness.  At fixed mass, a
more compact or steeper-profile dark perturber can produce a stronger lensing
signal than a more diffuse one
\citep{dalal_direct_2002,Vegetti_2009,Vegetti_2014,ritondale_lowmass_2019,VegettiReview2024,Despali_2025}.
Thus the relevance of the mass-segregated FCT subset is not that strong
lensing specifically requires a central PBH, but that PBH mass segregation
provides a physically motivated route to especially compact dark perturbers
within an already enhanced FCT subhalo population
\citep{Colazo_2025}.

This compact dark-substructure population is not a generic prediction of all
PBH scenarios.  It requires an extended PBH mass function with a high-mass
tail capable of sinking efficiently within low-mass haloes.  In the FCT model,
the same PBH discreteness that contributes to the modified initial power
spectrum also supplies such a high-mass tail.  The result differs from the
usual expectation in models with suppressed small-scale structure, such as
warm dark matter
\citep{Enzi_2021,anau_montel_2023,he_extending_2023,gilman_2024}.  In those
models the main expectation is a reduction in the number of low-mass
perturbers.  PBH/FCT models can instead produce an enhanced low-mass subhalo
abundance, with a small tail of subhaloes in which unresolved PBH mass
segregation adds a sizeable inner perturbation to the FCT reference profile.

Whether these mass-segregated substructures are luminous or dark is ultimately
a question of star-formation efficiency in low-mass haloes.  The selected
population in Fig.~\ref{fig:pdense_mass_function_highhosts} lies in the
low-mass regime where UV photoheating suppresses gas retention and where the
satellite occupation fraction is expected to become stochastic
\citep{gnedin2000,okamoto2008,Simon2019,Nadler2020}.  Massive PBHs in the
sunk component could provide an additional local heating channel in such
marginal systems, but quantifying this requires gas physics and accretion
modelling beyond the dark-matter-only simulations used here.  We therefore
treat the luminous occupation of this population as uncertain.

The subgrid calculation identifies a useful candidate population for future
lensing work: FCT substructures with enhanced low-mass abundance, high
inferred compactness, and a  PBH mass-segregation perturbation large
enough to be dynamically relevant relative to the FCT reference enclosed mass
at \(r_{\rm eff}\).  A direct strong-lensing prediction would require
projected density profiles for these objects, the full line-of-sight halo
population, and the lens and source selection functions.  These substructures
are therefore natural targets for follow-up calculations of lensing
perturbations and other gravitational signatures of compact dark
substructure.

\section{Discussion and conclusions}
\label{sec:discussion_conclusions}

We have studied how a fixed-conformal-time PBH model modifies satellite
subhalo abundances and internal structure relative to CDM.  Because the
simulations are dark-matter-only, we assign stellar masses using an
abundance-based correction anchored to the CDM stellar-to-halo mass relation
at infall.

The FCT model changes the shape, rather than only the normalization, of the
cumulative subhalo mass function.  The resulting stellar-to-halo correction
is therefore mass dependent.  After applying this correction, the CDM and FCT
satellite stellar-mass functions are similar over the well-sampled ranges of
Milky-Way and low-mass group hosts, with typical offsets
\(-0.15<\Delta\log_{10}M_\star\lesssim0\) in the GAMA-like comparison.  The
largest differences occur at low stellar masses, where satellite occupation
and observational incompleteness are most uncertain.  Satellite counts alone
are consequently a weak discriminator between the models.

The internal-structure comparison provides a stronger constraint.  FCT
subhaloes are more compact than CDM in the resolved SOAP diagnostics and
produce larger NFW-reconstructed enclosed masses at fixed radius.  If all of
these compact objects are assumed to host luminous satellites, the result
strengthens a Too-Big-To-Fail-like tension.  These estimates are
NFW-equivalent reconstructions based on resolved \(V_{\max}\) and \(R_{\max}\).

FCT subhaloes form substantially earlier than their CDM counterparts at low
infall mass. Under the approximate scaling \(c\propto1+z_{\rm form}\),
assigning FCT-like formation epochs to a CDM reference population accounts
for much of the low-mass concentration offset, but not for the difference at
higher masses. The subsequent subgrid PBH mass-segregation correction adds a
further central enhancement: the expected sunk PBH mass is typically an order-half perturbation relative
to the enclosed mass of the CDM-like reference profile over the subhalo-mass
range shown.

The FCT-reference calculation provides a stricter comparison because it
measures the sunk PBH component relative to an already compact FCT profile.
In massive hosts, systems with \(\mu_{\rm FCT}>0.5\) form a small subset in
which PBH segregation provides a sizeable additional inner perturbation.
If these low-mass systems form stars inefficiently, they may be more relevant
as compact dark perturbers than as luminous satellites.

Together, the formation-time and PBH-segregation corrections raise the
NFW-equivalent concentration well above the CDM baseline, while generally
remaining below the full FCT compactness proxy. Earlier assembly and
unresolved PBH segregation can therefore explain part of the FCT structural
enhancement, with the residual likely reflecting the nonlinear evolution of
the modified initial density field, subsequent tidal evolution, and the
limitations of the adopted formation-time scaling.

Our conclusions are limited by the small simulation volume, the approximate
abundance-matching prescription, observational incompleteness, and the
absence of baryonic disruption and a probabilistic galaxy--subhalo
occupation model.  The PBH sinking calculation is likewise a subgrid
estimate based on Chandrasekhar dynamical friction and does not model
individual PBH orbits or stochastic relaxation in a granular halo.

The quantitative results obtained here are not generic predictions
of all PBH scenarios. The enhancement of small-scale structure arises from
the particular modification of the initial power spectrum in the FCT model,
including the PBH discreteness contribution, and its magnitude depends on the
PBH abundance and mass function. The formalism of \citet{Padilla_2021} provides
a route to generalizing this calculation: for a given PBH abundance and mass
function, the corresponding modification of the initial matter power spectrum
can be constructed and compared with that of the FCT model. For extended mass
functions, the discreteness contribution depends on moments of the PBH mass
distribution and can therefore be particularly sensitive to its high-mass
tail. Likewise, efficient mass segregation requires sufficiently massive
PBHs, so the central enhancement estimated here also depends strongly on this
high-mass tail. More generally, PBH scenarios that generate comparable
additional small-scale power may produce qualitatively similar changes in
low-mass halo abundances and assembly histories. The nonlinear response of
the subhalo population, however, as well as the importance of PBH sinking,
must ultimately be evaluated for each PBH model separately.

In summary, satellite stellar-mass functions can partly absorb the modified
FCT subhalo abundance through a mass-dependent stellar-to-halo mapping,
whereas internal structure retains a clearer signature.  The earlier formation of FCT subhaloes increases their expected compactness,
while the subsequent segregation of massive PBHs can provide an additional
inner mass perturbation, although neither effect fully explains the resolved
FCT structural offset.  Satellite abundances,
satellite dynamics, and gravitational searches for compact dark perturbers
therefore provide complementary probes of PBH-induced modifications to
small-scale structure.

\begin{acknowledgements}
NP acknowledges support from PICT Raices Federal 2023-0002. 
\end{acknowledgements}

\bibliographystyle{aa}
\bibliography{bibliography}
\end{document}